\documentclass[conference]{IEEEtran}
\IEEEoverridecommandlockouts
\usepackage{cite}
\usepackage{amsmath,amssymb,amsfonts}
\usepackage{algorithmic}
\usepackage{graphicx}
\usepackage{textcomp}
\usepackage{subfig}
\usepackage{multirow}
\usepackage[multiple]{footmisc}
\usepackage{xcolor}
\usepackage{booktabs}
\usepackage{balance}
\usepackage{threeparttable}
\usepackage{tabularx}
\def\BibTeX{{\rm B\kern-.05em{\sc i\kern-.025em b}\kern-.08em
    T\kern-.1667em\lower.7ex\hbox{E}\kern-.125emX}}

\usepackage{tikz}
\usepackage[hyperfootnotes=false]{hyperref} 

\newcommand\copyrighttext{%
  \footnotesize \textcopyright 2021 IEEE. Personal use of this material is permitted.
  Permission from IEEE must be obtained for all other uses, in any current or future
  media, including reprinting/republishing this material for advertising or promotional
  purposes, creating new collective works, for resale or redistribution to servers or
  lists, or reuse of any copyrighted component of this work in other works.
  DOI: \href{https://doi.org/10.1109/BigData52589.2021.9671275}{https://doi.org/10.1109/BigData52589.2021.9671275}}
\newcommand\copyrightnotice{%
\begin{tikzpicture}[remember picture,overlay]
\node[anchor=south,yshift=10pt] at (current page.south) {\fbox{\parbox{\dimexpr\textwidth-\fboxsep-\fboxrule\relax}{\copyrighttext}}};
\end{tikzpicture}%
}

\DeclareMathAlphabet{\altmathcal}{OMS}{cmsy}{m}{n}
    
\begin{document}

\title{On the Potential of Execution Traces for Batch Processing Workload Optimization in Public Clouds}

\author{
\IEEEauthorblockN{Dominik Scheinert\IEEEauthorrefmark{1}, Alireza Alamgiralem\IEEEauthorrefmark{1}, Jonathan Bader\IEEEauthorrefmark{1}, \\Jonathan Will\IEEEauthorrefmark{1}, Thorsten Wittkopp\IEEEauthorrefmark{1}, and Lauritz Thamsen\IEEEauthorrefmark{3}\IEEEauthorrefmark{1}}
\IEEEauthorblockA{\IEEEauthorrefmark{1}Technische Universit{\"a}t Berlin, Germany, \{firstname.lastname\}@tu-berlin.de}
\IEEEauthorblockA{\IEEEauthorrefmark{3}Humboldt-Universit{\"a}t zu Berlin, Germany, lauritz.thamsen@hu-berlin.de}
}

\maketitle
\copyrightnotice

\begin{abstract}
With the growing amount of data, data processing workloads and the management of their resource usage becomes increasingly important.
Since managing a dedicated infrastructure is in many situations infeasible or uneconomical, users progressively execute their respective workloads in the cloud. 
As the configuration of workloads and resources is often challenging, various methods have been proposed that either quickly profile towards a good configuration or determine one based on data from previous runs. 
Still, performance data to train such methods is often lacking and must be costly collected.

In this paper, we propose a collaborative approach for sharing anonymized workload execution traces among users, mining them for general patterns, and exploiting clusters of historical workloads for future optimizations.
We evaluate our prototype implementation for mining workload execution graphs on a publicly available trace dataset and demonstrate the predictive value of workload clusters determined through traces only.
\end{abstract}


\section{Introduction}
\label{sec:introduction}

Large cloud providers offer a wide range of virtual resources and managed services, and thus serve various use cases and needs.
From a user perspective, this has various advantages, such as dynamic scaling of resources based on demand, or simplified management and deployment.
Especially for infrequent needs, making use of cloud resources is often favored over running and managing own resources.
Hence, an increasing number of computing workloads are run in the cloud.

Yet, in the majority of cases, users still need to select appropriate resources for their workloads.
Inexperienced users are often overwhelmed by the range of different resources to choose from, while even expert users find this to be difficult.
With an increasing number of scientists and data analysts from other domains that require data-processing workloads to be run in the cloud~\cite{Leser_2013,DeelmanVRMSPL19,bader2021tarema}, supporting them becomes important.

Multiple solutions have been proposed over the years to support users with fixed constraints, e.g. in terms of runtime or cost.
They can be roughly clustered into search-based methods~\cite{AlipourfardLCVY17,HsuNMF18,0007CR20} and model-based methods~\cite{VenkataramanYFR16,ShahAKW19,ScheinertTZWAWK21,will2021c3o}.
In general, most methods require historical data about workload executions or benefit from them, yet their collection is costly or often restricted to certain use cases. 
However, since cloud users usually operate on the same infrastructure and possibly execute similar workloads, there lies potential in sharing and exploiting information about their workload executions, most likely in the form of confidentiality-preserving traces of data. 

In this paper, we envision a system for collaborative exploitation of workload execution traces as well as data enrichment in order to optimize future workloads.
We further argue for an explicit consideration of the attributed graph of task-dependencies of workloads, and strive for computing compact representations of these graphs for downstream clustering tasks. 
Ideally, computed clusters consist of similar or reoccurring workloads. 
Knowledge present in individual clusters, e.g. average runtime of workloads, can then be used to make better decisions for new workloads.
Our approach to encoding and clustering of workload execution graphs is prototypically evaluated on the Alibaba dataset\footnote{\href{https://github.com/alibaba/clusterdata}{https://github.com/alibaba/clusterdata}} of cluster traces. 
We are able to demonstrate that even anonymized workload execution graphs bear a predictive value for identifying groups of similar workloads. 
This underlines the potential of exploiting and enriching shared workload execution traces for predicting performance indicators of workloads, and addresses the problem of limited data via data sharing.

\emph{Contributions}. The contributions of this paper are:
\begin{itemize}
    \item An idea for a system that enables users to share traces of their workload executions. 
    Exchanged data can be evaluated for general patterns and workload similarities, which can be exploited for optimization of future workloads.
    \item A prototypical implementation for mining of traces from workload execution graphs via graph encoding and graph clustering. 
    These steps form an important part of the optimization process.
    \item An evaluation of our implementation on a publicly available trace dataset. We demonstrate the predictive value of workload execution graph traces, e.g. for finding clusters of presumably similar workloads, and discuss the implications of our findings.
\end{itemize}

\emph{Outline}. 
\autoref{sec:idea} elaborates on the idea and proposes a system for exploiting workload execution traces, whereas \autoref{sec:approach} concretizes on the encoding and clustering of workload execution graphs. 
\autoref{sec:results} presents the preliminary results of our trace analysis, and is followed by a discussion of general requirements of our approach in \autoref{sec:discussion}.
\autoref{sec:related_work} discusses the related work.
\autoref{sec:conclusion} concludes the paper.

\section{System Idea}
\label{sec:idea}
\begin{figure}[h!]
    \centering
    \includegraphics[width=\columnwidth]{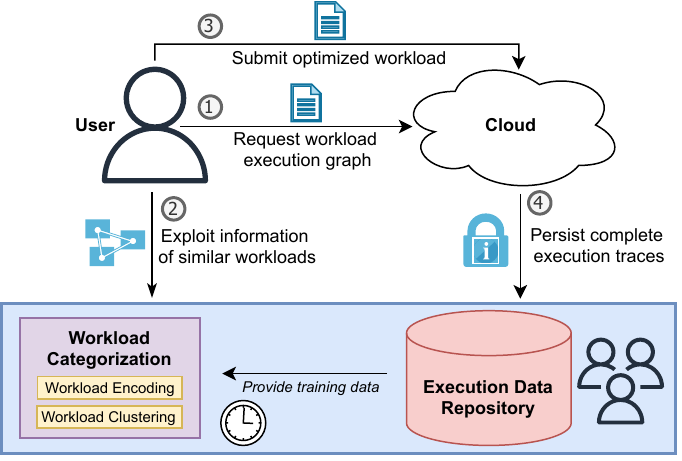}
    \caption{Overview of envisioned system}
    \label{fig:general_idea}
\end{figure}

This section elaborates on our envisioned system for workload optimization through 
data sharing and exploitation of workload traces.
It is further illustrated in~\autoref{fig:general_idea}.

\subsection{Sharing of Confidential Execution Data}
Workload optimization through selection of more suitable cloud configurations is often realized with performance models. 
However, such models usually require a certain amount of historical data or profiling, which is costly to collect.
Yet, many cloud users operate on the same infrastructure and potentially even run similar workloads, which opens the possibility of data sharing.
As execution data of workloads encompasses also confidential and personal information, it is necessary to share traces of this data in a confidentiality-preserving manner.
In the context of recent publications of cluster trace datasets
from cloud providers, research was also conducted on this aspect~\cite{ReissWH12}.
If cloud providers were to implement interfaces for conveniently fetching execution traces, or if enough users organize themselves and gather their respective traces in a centralized place, methods can be employed for exploiting data and optimizing future workloads.

\subsection{Envisioned System}
Assuming that for each planned or executed workload, its execution graph traces are accessible, cloud users can collect their traces, optionally enrich them with further information describing the respective workloads, and save them to community-managed execution data repositories. 
With growing data, machine learning (ML) methods can be employed to encode and cluster the various workloads according to certain characteristics and performance indicators.
As the traces are confidential by design, and users add further data on a voluntarily basis, users retain full control over their data.

Whenever a user attempts to submit a new workload, traces of the prospective execution plan are retrieved from the respective cloud provider at first.
The information can be used to consult the various trained ML methods in order to exploit insights of similar historical workloads. 
With this knowledge gain, the desired workload can be optimized according to user-specific objectives and constraints, and eventually submitted.

\section{Mining of Execution Graph Traces}
\label{sec:approach}
\begin{figure}[h!]
    \centering
    \includegraphics[width=.8\columnwidth]{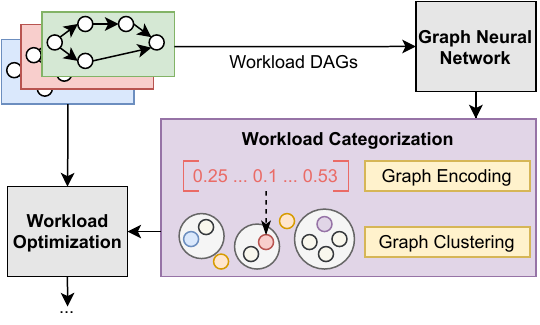}
    \caption{The proposed graph processing pipeline}
    \label{fig:approach}
\end{figure}

This section presents our approach to the mining of execution graph traces of workloads for the optimization of future workloads.
It encompasses both encoding and clustering steps.
The complete approach is generally sketched in~\autoref{fig:approach}.

\subsection{Preliminaries}
Data-parallel processing jobs can be modelled as directed acyclic graphs (DAG) to represent the dependencies between job tasks.
A directed, acyclic, and attributed graph $G=(V,E)$ consists of a set of $n$ nodes $V=\{v_1, \ldots, v_n\}$ and a set of edges $E\subseteq \{(v_i,v_j)| v_i,v_j \in V\}$. 
Each node $v_i$ has a node feature vector $\Vec{x}_i \in \mathbb{R}^{F}$ of dimension $F$.
An edge $e_{ij} \Leftrightarrow (v_i,v_j)\in E$ describes a directed connection between node $v_i$ and $v_j$. 
Thus, the node $v_j$ is then called a neighbor of node $v_i$, formally written as $j\in \altmathcal{N}(i)$. 
The adjacency matrix $A$ of a graph $G$ is an $n \times n$ matrix with entries $A_{ij}$ such that $A_{ij}=1$ if an edge $e_{ij}$ exists, else 0.
No cycles exist in the graph.

\subsection{Graph Encoding}
In order to leverage the DAG of a workload in various prediction models, we need to compute a reasonable representation in the first place.
Thus, we compose a graph neural network (GNN) architecture and derive a vector representation for each graph through a final global pooling operation.
In a first step, multiple graph convolutional layers are employed.
The fundamental idea of our GNN is the exploitation of structural information in the graph, which is realized by passing node features to neighboring nodes and thereby constantly computing new node features. 
This operation is referred to as \emph{neighborhood aggregation} and commonly defined as

\begin{equation}
    \Vec{x}_i^{(k)} = \gamma^{(k)} \Big( \Vec{x}_i^{(k-1)}, \lambda_{j\in \altmathcal{N}(i)} \phi^{(k)} \Big( \Vec{x}_i^{(k-1)}, \Vec{x}_j^{(k-1)}, e_{ji} \Big) \Big),
\end{equation}

where $\lambda$ denotes a differentiable and permutation invariant function, $k$ denotes the number of hops, and both $\gamma$ and $\phi$ denote differentiable functions, e.g. feed-forward neural networks, which optionally alter the vector dimensionality.
When conducting multiple such steps, structural information are effectively exploited and propagated through the graph. 

Eventually, we seek to compute a fixed-size vector representation for a graph. 
This can be formally expressed as

\begin{equation}
\Vec{g}(i) = \lambda_{j\in E(G)}\Vec{x}_j(i),
\end{equation}

where $\Vec{g}\in \mathbb{R}^M$ is the final vector representation for a graph $G$ with hidden dimension $M$, and $\Vec{x}_j(\cdot)$ is used to access individual elements in the feature vector of the $j$-th node.
Since $\Vec{g}$ shall be a good representation of the corresponding original graph $G$, it is necessary to design the training of the graph neural network appropriately. 
Given a set of graph features, e.g. number of nodes or the length of the longest path, chosen via a feature selection process based on their predictive value for some target objective, we argue for employing an additive loss.
With each loss term measuring how good computed graph representations can be separated given a concrete graph feature, the optimization process eventually leads to the generation of graph representations which capture the most relevant information of the original graph.
As all representation vectors maintain a suitable dimensionality, they can be used for downstream prediction and planning tasks.

\subsection{Graph Clustering}
With the computed graph representations containing as much original information as possible, they can be further clustered into groups of vectors which are presumably similar in terms of their structure and node annotations. 
This is also a meaningful indicator for identifying recurring or mostly similar batch workloads.
In order to make use of clustering techniques, the graph representations first need to be preprocessed. 
We standardize features by removing the mean and scaling to unit variance, and then scale the graph representations individually to unit norm. 
The preprocessed vector of a graph $G$ is from here on denoted as $\Vec{g'}$.

To determine the distance between two graph representations, we compute their Euclidean distance
\begin{equation}
    d(\Vec{g'}_i, \Vec{g'}_j) = \sqrt{\sum_{k=1}^M \Big(\Vec{g'}_i(k) - \Vec{g'}_j(k)\Big)^2},
\end{equation}
where $\Vec{g'}_i$ and $\Vec{g'}_j$ denote representations of two different graphs respectively.
Computed vector distances are then utilized during clustering.
As the number of clusters evolves over time and thus can not be known a priori, we utilize a density-based clustering technique. 
Suitable hyperparameters for such a technique can be determined empirically during training. 

\subsection{Information Enrichment \& Optimization}
The identified clusters of data points are a first indication of executions of either identical or similar batch workloads. 
Yet, as more concrete information about the nature of a workload are not present in most cluster traces, either due to cluster providers not being aware of e.g. input dataset characteristics or not being allowed to publish confidential and personal information, an even more precise clustering solely based on cluster traces can not be achieved.
Following our envisioned system, users are ideally able to obtain traces of their own workload executions and voluntarily enrich them with exclusive information about utilized datasets, models, or workload results. 
In such a scenario, the interpretation of data point clusters would be simplified.

The clustering results can be used to exploit properties common to the majority of data points belonging to a concrete cluster.
Depending on the overall objective to optimize, information can be leveraged for future workloads.
For example, assuming that the graph representation learning has been optimized for the task of runtime prediction and that clusters of data points have been determined based on the resulting graph representations, the execution graph of a new workload, prior to its actual execution, could be used to identify the most similar cluster and compute the median runtime of its corresponding data points. 
This runtime estimate can then be used for workload optimization, e.g. by tweaking the associated resource configuration of the workload.

\section{Preliminary Trace Analysis}
\label{sec:results}
In this section, we examine the predictive value of workload DAGs using a prototype implementation of our presented approach called \emph{Trace-EC}. 
We make use of a comprehensive and publicly available cluster trace dataset and obtain preliminary results. 
Since, to the best of our knowledge, current cluster providers do not offer the possibility of independently exporting cluster traces of own workloads, and thus no investigation in conjunction with additional properties is possible, we solely assess the value of our approach based on the DAGs.

\subsection{Dataset}
The Alibaba cluster trace dataset\footnote{\href{https://github.com/alibaba/clusterdata}{https://github.com/alibaba/clusterdata}} consists of trace data of about 4000 machines over a period of 8 days. The required DAG information of workloads, also referred to as \emph{jobs}, are retrievable from the \emph{batch\_task} table. This table includes more than 14 million tasks of more than 4 million jobs and has eight features, namely \emph{task name}, \emph{job name}, \emph{start time} and \emph{end time}, \emph{status}, \emph{planned CPU} (100 means one core, 200 means two cores, etc.) and \emph{planned memory} (normalized to range $[0,100]$), as well as \emph{instance number} (the number of instances required for each task).
Since task dependencies are encoded into the task name, these features can be used to construct valid and attributed workload DAGs.

As the dataset suffers from minor inconsistencies and incompleteness, we employ multiple data cleaning operations. 
For instance, we exclude all unfinished jobs or jobs with undefined values. 
We furthermore make sure that the graph of each job complies with the formal definition of DAGs. 
For the sake of our evaluation, we eventually consider only the jobs that have at least 10 individual tasks to avoid coincidentally similarity in job dependencies structure. 
We also eliminate the jobs with runtime of longer than one hour.
The total number of remaining jobs is $232.809$ after data cleaning.
For each job, we further extract 20 features from its DAG (e.g. number of nodes) and consider them as candidates for target variables during model training.
Each node in a DAG has three node features, i.e. the aforementioned features planned CPU, planned memory, and instance number.
Using continuous data stratification based on the runtime variable, we split the dataset into 80\% training data and 20\% test data.

\subsection{Prototype Pipeline}
We implement our envisioned pipeline in a prototypical manner using various established methods.
In a first step, we employ a voting mechanism of multiple machine learning models and techniques to select the most runtime predictive features.
Precisely, we use Principal Component Analysis (PCA), Extra-trees, and Recursive Feature Elimination (RFE) with linear regression, and choose the union of the five top scored features of each technique to extract the target variables. 
They encompass for instance the \emph{average node degree} and \emph{total instance number}.
For supervised techniques, \emph{runtime} is defined as the target variable and excluded from the features.

The graph neural network is implemented using four stacked \emph{MFConv}~\cite{DuvenaudMABHAA15} layers, where the first one maps from the node feature dimension of 3 to the hidden dimension of 64.
Each graph convolution layer is followed by a non-saturating ELU activation. The final node representations are averaged graph-wise to obtain graph representations.
For regularization during training, we apply dropout with 50\% probability on the model output, as well as an L2 penalty term of $10^{-4}$ for weight decay. 

Using continuous data stratification based on the runtime variable, the provided graph training dataset is split into 75\% training data and 25\% validation data.
The latter is used for early stopping.
We use the Adam optimizer and a batch size of 128 graphs during training. 
We decide for an additive loss to be minimized, where each loss term is a \emph{triplet loss}~\cite{SchroffKP15} term corresponding to an individual target variable.
Triplet loss reduces the distance between similar objects and increases the distance between dissimilar ones, based on a target variable.
With a \emph{MultiSimilarityMiner}~\cite{WangHHDS19}, we make sure to evaluate challenging triplets in the respective triplet loss terms.

For the clustering of graph representations, we employ DBSCAN~\cite{EsterKSX96}, with radius $10^{-3.5}$ and a minimum of two samples in order to form a cluster.

\subsection{Baseline}
In~\cite{TianZ019}, the authors discuss a methodology for identifying recurrent jobs.
It is applied on the Alibaba trace dataset for gaining potential insights.
They argue that two or multiple jobs could be assumed as recurrent when they fulfill two conditions:
\begin{enumerate}
  \item The jobs have isomorphic DAGs.
  \item The start time of jobs should happen within periodical time intervals, e.g. 15 minutes, 1 hour, or 1 day intervals.
\end{enumerate}
Both conditions were individually shown to be relevant in related works, for instance in scheduler design or for planning of production jobs.
We thus utilize this methodology to compute clusters and compare them against our envisioned approach. 
\emph{Bliss}~\cite{JunttilaK07} is used for extracting isomorphic graphs, and the aforementioned time intervals (relatively expanded for tolerance by adding 3\% of intervals before and after) are investigated for deriving clusters of jobs. 

\subsection {Evaluation}
Since the utilized test dataset of 46.562 workload DAGs has no ground truth, i.e. no information about group affiliation of workloads exist, a precise evaluation is hindered.
Thus, we make the following approximation to interpret our results. 
Given a batch workload that is executed recurrently, it can be assumed that most executions exhibit similar characteristics and are thus strongly related.
The clusters identified by the baseline and our approach should reflect this at best, i.e. clustered executions should either originate from the same workload or be very similar. 
We thus assess the goodness of clusters in two ways.
In a first step, we test each clustering individually by predicting the runtime of the most recent execution in each cluster as the average runtime of all remaining cluster members. 
Consequently, this yields as many predicted values as clusters were formed.
The prediction errors are measured and reported in terms of Mean Absolute Error (MAE) and Mean Squared Error (MSE).
In a second step, we attempt to compare both clusterings.
Since the number of identified clusters differs and no groundtruth is given, we do so by extracting the subset of workloads predicted in both clusterings, and comparing the associated prediction results.

\subsection{Results}
\begin{table}
\centering
\begin{threeparttable}
    \caption{Comparison of method performance.}
    \begin{tabularx}{.8\columnwidth}{c||c|c}
    & \textbf{Baseline} & \textbf{Trace-EC}\\
    \toprule
    Proportion of outliers & 38,52\% & 39,44\%\\
    \#Clusters & \textbf{7.202} & \textbf{5.233}\\
    Avg. cluster size & \textbf{18,16} & \textbf{42,62}\\
    MAE & \textbf{143,33} & \textbf{84,55}\\
    MSE & 96.319,00 & 35.533,42\\
    Variance & 37.147,10 & 12.714,63\\
    \bottomrule
    \end{tabularx}
\label{tbl:evaluation_results}
\end{threeparttable}
\end{table}

\begin{figure*}
\centering
\subfloat{
  \includegraphics[width=\columnwidth]{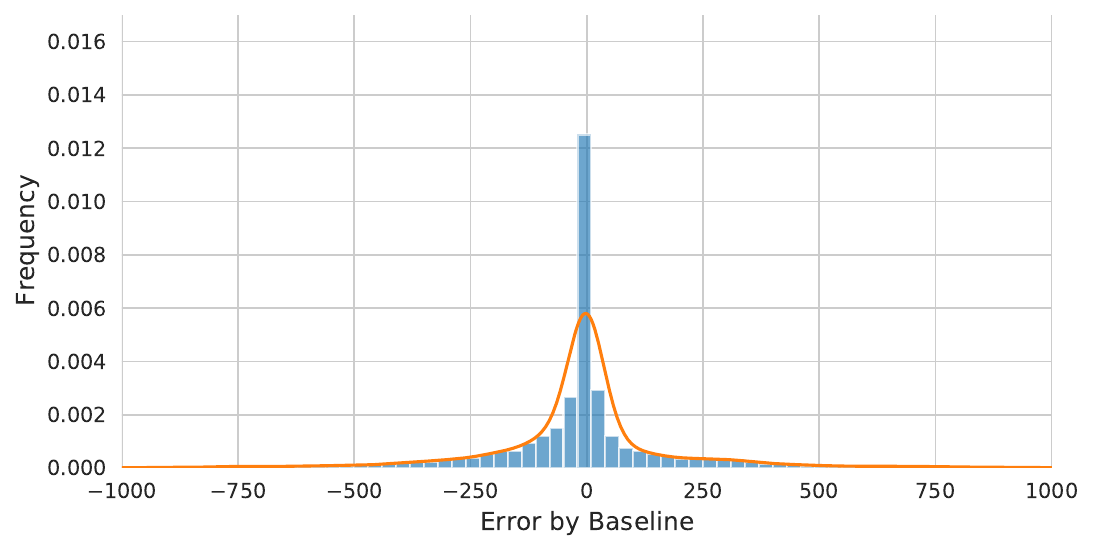}
}
\subfloat{
  \includegraphics[width=\columnwidth]{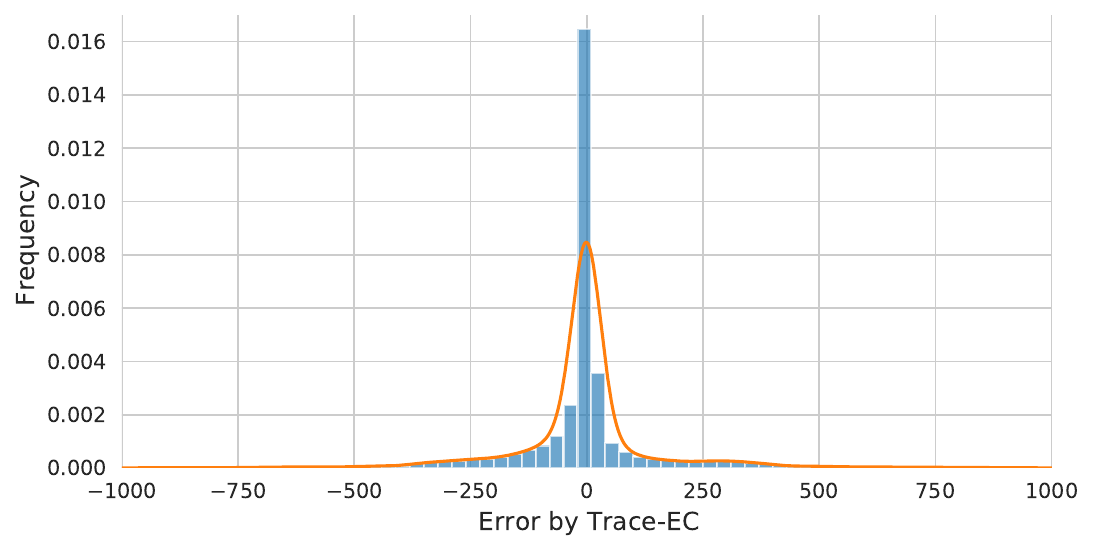}
}
\caption{Histograms of prediction errors and their direction for both Baseline and Trace-EC on clustered test data. The proportions of test samples falling into the interval [-1000, 1000] are 98,11\% and 99,48\% respectively. Trace-EC produces smaller errors.}
\label{fig:evaluation_individual}
\end{figure*}

\begin{figure}
    \centering
    \includegraphics[width=\columnwidth]{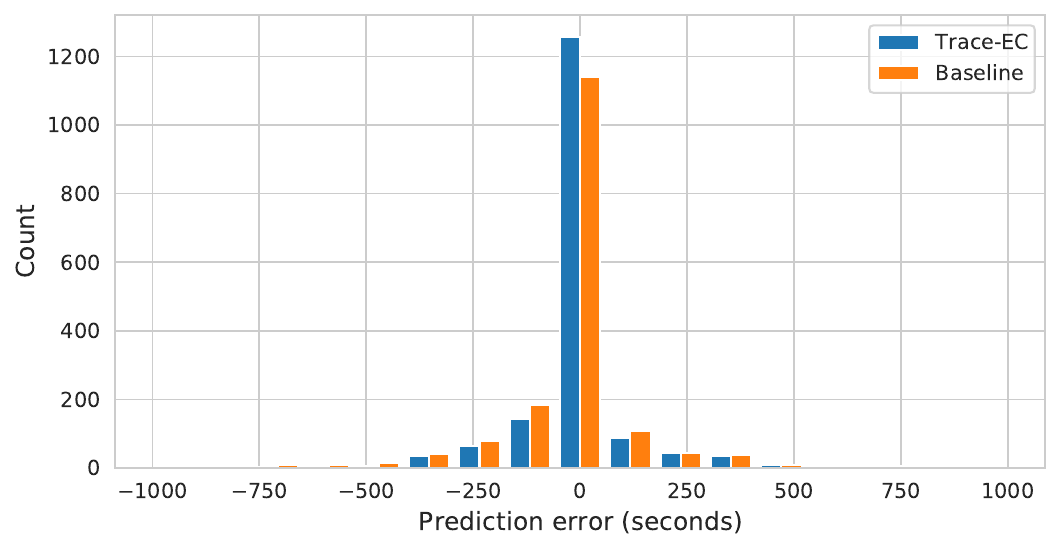}
    \caption{Direct method comparison using the subset of shared target workloads and their associated generated clusters. The clustering of Trace-EC allows for smaller prediction errors.}
    \label{fig:evaluation_comparison}
\end{figure}

\autoref{tbl:evaluation_results} summarizes the most important findings of our first evaluation step.
It can be seen that although both methods are able to cluster roughly the same amount of workloads, Trace-EC tends to form fewer clusters with more members each, whereas the utilized baseline suspects more clusters with on average less members.
Considering the composition of individual clusters, it is worth mentioning that Trace-EC apparently produces well-formed clusters, as the reported MAE is substantially smaller compared to the baseline.
This can be also found for the reported MSE and variance.
Furthermore, the histograms in~\autoref{fig:evaluation_individual} underline that clusters generated by Trace-EC allow for more accurate predictions and in turn smaller prediction errors.
This results are consequently a first indication of superior performance.

For our second evaluation step, we extract the subset of workloads predicted in both methods, which leaves us with 1715 clusters from both clusterings. 
Again, we compute the relevant evaluation metrics on the subsets and illustrate them in~\autoref{fig:evaluation_comparison}.
It is confirmed once more that Trace-EC yields better capabilities for downstream optimization tasks, which becomes evident by its MAE of 72,62 outperforming the MAE of 94,71 from the baseline.
In conclusion, clusters identified by Trace-EC tend to encompass more similar workloads, with respect to desired performance objectives.

\section{Discussion}
\label{sec:discussion}
Our findings demonstrate that the explicit consideration of workload DAGs through a flexible data-driven approach yields improved prediction performance. 
Even with normalized attributes, workload DAGs obtained from cluster traces appear to have predictive value.
Although the lack of ground truth hinders a complete evaluation, generated clusters tend to encompass similar workloads and can thus be exploited for future workload optimization.

For application in real-world cloud environments and for optimization of actual workloads, we deem two things important.
On the one hand, users are still required to enrich their collected traces with further information, e.g. dataset size or chosen algorithm implementation. 
Without such information, a more fine-grained distinction and thus clustering of workloads is hardly possible. 
On the other hand, cloud providers need to provide simple ways of retrieving confidential traces for own workloads. 
This can for instance be achieved by configurable data sinks, such that each workload execution is followed by an automated transfer of anonymized traces to a desired storage, e.g. a community-managed execution data repository.

\section{Related Work}
\label{sec:related_work}
Our work motivates the usage of cluster traces in order to collaboratively optimize workloads.
This section consequently discusses attempts to collaborative workload optimization, as well as other trace analysis papers and trace-based methods.

\subsection{Collaborative Workload Optimization} 
Multiple solutions have been proposed that either enable the usage of data originating from various contexts, strive to find a configuration that optimizes numerous workloads at the same time, or foster a collaborative approach in a different way.

Micky~\cite{HsuNMF18} is a collective optimizer that determines a cloud configuration optimizing as many of the given workloads as possible.
In this specific context, operational costs and execution times are regarded as performance objectives, and an optimized solution balances both of them for most workloads.

The Peregrine~\cite{JindalPRQYSK19} workload optimization platform follows a collaborative approach by searching for patterns in historical query workloads and employing suitable optimization strategies.
Exemplary patterns are periodicity and similarity.

The authors of~\cite{DerakhshanMARM20} attempt to optimize workloads in environments with multiple actors by utilization of an experiment graph that allows for reuse of historical operations and their artifacts. 
Their work is especially suited for machine learning workloads while incurring only negligible overhead. 

In our own previous work, we investigate collaborative solutions for distributed dataflows. 
Bellamy~\cite{ScheinertTZWAWK21} proposes an end-to-end trainable neural network architecture that allows for incorporation of workload execution data originating from different contexts. 
It can thus be pre-trained and later fine-tuned on new contexts with only little available data.
Another work is the C3O system~\cite{WillBT20,will2021c3o,will2021training}, which enables the sharing of runtime data and artifacts.
This allows for collaborative exploitation of shared information, which is demonstrated to be effective in case of context-aware predictors.

\subsection{Cluster Traces from Cloud Providers}
In recent years, various cloud providers published traces of their clusters for further analysis and application.
A considerable proportion of works mainly focuses on analyzing workload characteristics and distributions~\cite{ReissTGKK12,LuYXXB17,ChengCA18,GuoCWDFMB19,TirmaziBDHQHHW20}, where it is for instance found that the resource and memory consumption of most workloads entails a heavy-tailed distribution, i.e. few workloads consume the majority of resources~\cite{ReissTGKK12,TirmaziBDHQHHW20}. 
At the same time, especially long-running workloads tend towards over-provisioning of resources~\cite{LuYXXB17}.

With regards to application, publicly available cluster traces are furthermore used for proposing novel scheduling approaches~\cite{LiuSSC18,WuZXX0DZ19}, or utilizing trace information for implementing or evaluating various prediction methods~\cite{DiKC12,ElSayedZS17}.

After individual cloud providers started to include information about the DAG of workloads in their traces, researchers also began to investigate the cluster traces with regards to the graph-structured task dependencies of workloads~\cite{TianZ019,lu2020understanding}.
Our work is most similar to theirs.
\section{Conclusion}
\label{sec:conclusion}
The primary goal of this work is to show how even anonymized execution traces shared with other users can be used to optimize future workloads.
To this end, we envision a system that fosters trace data sharing among users, such that users can design and employ methods for detection of patterns, which in turn can be exploited for planning and optimizing workloads in the future, e.g. with regards to runtime or cost.
Towards this goal, we implemented an approach for encoding and clustering traces of workload execution graphs, and evaluated it on a publicly available trace dataset.
We find that our data-driven solution is able to make use of the predictive value of workload DAGs for performance indicators of interest, which in turn is a prerequisite for workload optimization.

In the future, we plan to leverage our findings and use graph information for optimized resource management for data processing workloads, including for resource allocation and scheduling. Moreover, we want to investigate the potential of traces other than execution graphs.

\section*{Acknowledgments}
This work has been supported through grants by the German Federal Ministry of Education and Research (BMBF) as BIFOLD (funding mark 01IS18025A) and the German Research Foundation (DFG) as FONDA (DFG Collaborative Research Center 1404).

\bibliographystyle{IEEEtran}
\balance
\bibliography{bib}

\end{document}